# Analytic Representation of
# The Dirac Equation

Tepper L. Gill[1,2,3] and W. W. Zachary[1,2]

[1]Department of Electrical and Computer Engineering
[2]Computational Physics Laboratory
[3]Department of Mathematics
Howard University
Washington, D. C. 20059

**Abstract**

In this paper we construct an analytical separation (diagonalization) of the full (minimal coupling) Dirac equation into particle and antiparticle components.  The diagonalization is analytic in that it is achieved without transforming the wave functions, as is done by the Foldy-Wouthuysen method, and reveals the nonlocal time behavior of the particle-antiparticle relationship.   We interpret the zitterbewegung and the result that a velocity measurement (of a Dirac particle) at any instant in time is $\pm c$, as reflections of the fact that the Dirac equation makes a spatially extended particle appear as a point in the present by forcing it to oscillate between the past and future at speed $c$.  From this we infer that, although the form of the Dirac equation serves to make space and time appear on an equal footing mathematically, it is clear that they are still not on an equal footing from a physical point of view. On the other hand, the Foldy-Wouthuysen transformation, which connects the Dirac and square root operator, is unitary.  Reflection on these results suggests that a more refined notion (than that of unitary equivalence) may be required for physical systems.
   We then show explicitly that the Pauli equation is not completely valid for the study of the Dirac hydrogen atom problem in s-states (hyperfine splitting).  We conclude that there are some open mathematical problems with any attempt to explicitly show that the Dirac equation is insufficient to explain the full hydrogen spectrum. Our analysis suggests that the use of cutoffs in QED is already justified by the eigenvalue analysis that supports it if the perturbation method can be justified.
   Using a new method, we are able to effect separation of variables for full coupling and solve the radial equation.  The behavior of the radial equation at the origin is the same as in the Dirac-Coulomb case, so that the $\mathbf{A}^2$ term, which appears in an exact analysis, does not increase the singular nature of the wave function.

**I. Introduction**

It is generally agreed that quantum electrodynamics (QED) is an almost perfect theory that is in excellent agreement with experiments. The fact that it is very successful is without doubt. However, there are still some technical and foundational issues which require clarification.

Historically, when Lamb and Retherford[1] confirmed suspicions that the $2s_{1/2}$ state hydrogen was shifted above the $2p_{1/2}$ state, the Pauli approximation to the Dirac equation was (essentially) used to decide that the Dirac equation was not sufficient.

**II Purpose**

In light of the tremendous success (historically) of eigenvalue analysis in physics and engineering, it is not inappropriate for us to reinvestigate the foundations of spin 1/2 particles with an eye towards clearly identifying the conceptual, physical and mathematical limitations to our understanding of the hydrogen spectrum as an eigenvalue problem.

The first successful attempt to resolve the question of how best to handle the square-root equation:

$$i\hbar \frac{\partial \Psi}{\partial t} = \beta \left[ \sqrt{c^2 \mathbf{p}^2 + m^2 c^4} \right] \Psi, \quad \beta = \begin{bmatrix} I_2 & 0 \\ 0 & -I_2 \end{bmatrix}, \quad (1)$$

was made by Dirac[2] in 1926. Dirac noticed that the Pauli matrices could be used to write $c^2 \mathbf{p}^2 + m^2 c^4$ as $[c\boldsymbol{\alpha} \cdot \mathbf{p} + mc^2 \boldsymbol{\beta}]^2$. The matrix $\boldsymbol{\alpha}$ is defined (in the standard representation) by $\boldsymbol{\alpha} = (\alpha_1, \alpha_2, \alpha_3)$, where

$$\alpha_i = \begin{pmatrix} \mathbf{0} & \sigma_i \\ \sigma_i & \mathbf{0} \end{pmatrix}, \quad \sigma_1 = \begin{pmatrix} 0 & 1 \\ 1 & 0 \end{pmatrix}, \quad \sigma_2 = \begin{pmatrix} 0 & -i \\ i & 0 \end{pmatrix}, \quad \sigma_3 = \begin{pmatrix} 1 & 0 \\ 0 & -1 \end{pmatrix}.$$

Thus, Dirac showed that an alternative representation of equation (1) could be taken as:

$$i\hbar \frac{\partial \Psi}{\partial t} = [c\boldsymbol{\alpha} \cdot \mathbf{p} + mc^2 \boldsymbol{\beta}] \Psi. \quad (2)$$

In this case, $\Psi$ must be viewed as a vector-valued function or spinor. To be more precise, $\Psi \in L^2(\mathbf{R}^3, \mathbf{C}^4) = L^2(\mathbf{R}^3) \otimes \mathbf{C}^4$ is a four-component column vector $\Psi = (\psi_1, \psi_2, \varphi_1, \varphi_2)^t$. In this approach, $\psi = (\psi_1, \psi_2)^t$ represents the particle (positive energy) component, and $\varphi = (\varphi_1, \varphi_2)^t$ represents the antiparticle (negative energy) component of the theory (for details, see Thaller[3]).

A fair understanding of the Dirac equation can only be claimed in recent times, and, as pointed out by D. Finkelstein, "Dirac introduced a Lorentz-invariant Clifford algebra into the complex algebra of observables of the electron". (See, in particular, Biedenharn[4] or deVries[5] along with the references therein. For a general reference to Clifford algebras, the work by Hestenes[6] offers a good introduction.)

Despite successes, both practical and theoretical, there still remain a number of conceptual, interpretational, and technical misunderstandings about this equation. It is generally believed that it is not possible to separate the particle and antiparticle components directly without approximations (when interactions are present). The various approximations found in the literature might have led to this belief. In addition, the historically important algebraic approaches of Foldy-Wouthuysen[7], Pauli[8], and Feynman and Gell-Mann[9] have no doubt further supported such ideas.

We show in Section III that it is possible to directly separate the particle and antiparticle components of the Dirac equation without approximations, even when scalar and vector potentials of quite general character are present. In Section IV, we show that the square root operator cannot be considered physically equivalent to the Dirac operator. In addition, we offer another interpretation of the zitterbewegung and the fact that expected value of a velocity measurement of a Dirac particle at any instant of time is $\pm c$. In Section V, we reconsider the hydrogen atom problem from an exact point of view and then discuss the extent that we may believe in the validity of the use of perturbation analysis to compute the hyperfine splitting separation. Finally, in Section VI we show, using a combination of methods developed by Harish-Chandra and Villalba, that it is possible to effect a separation of variables for the hydrogen atom problem with the magnetic dipole vector potential. This allows us to provide some justification for perturbation analysis (to compute the hyperfine splitting) in this case.

**II. Complete Separation**

It turns out that a direct analytic separation is actually quite simple and provides additional insight into the particle and antiparticle components. In order to see this, let $\mathbf{A}(\mathbf{x},t)$ and $V(\mathbf{x})$ be given vector and scalar potentials and, after adding $V(\mathbf{x})$ and making the transformation $\mathbf{p} \to \boldsymbol{\pi} = \mathbf{p} - (e/c)\mathbf{A}$, write (2) in two-component form as:

$$i\hbar \frac{\partial \psi}{\partial t} = (V + mc^2)\psi + c(\sigma \cdot \boldsymbol{\pi})\varphi, \tag{3a}$$

$$i\hbar \frac{\partial \varphi}{\partial t} = (V - mc^2)\varphi + c(\sigma \cdot \boldsymbol{\pi})\psi. \tag{3b}$$

Equation (3b) can be written in the form:

$$\left[\frac{\partial}{\partial t} + iB\right]\varphi = D\psi, \qquad (4)$$

with $B = \left[(V - mc^2)/\hbar\right]$ and $D = \left[c(\sigma \cdot \pi)/i\hbar\right]$. From an analytical point of view, we see that equation (4) is an inhomogeneous partial differential equation. This equation can be solved via the Green's function method. Thus, we then must solve

$$\left[\frac{\partial}{\partial t} + iB\right]\varphi = \delta(t). \qquad (5)$$

It is easy to see that the solution to equation (5) is

$$u(t) = \theta(t)\exp\{-iBt\}, \quad \theta(t) = \begin{cases} 1, & t > 0 \\ 0, & t < 0 \end{cases}, \qquad (6)$$

so that,

$$\varphi(t) = cu(t) * \left[(\sigma \cdot \pi)/i\hbar\right]\psi(t) = \int_{-\infty}^{\infty} c\theta(t-\tau)\exp\{-iB(t-\tau)\}\left[(\sigma \cdot \pi)/i\hbar\right]\psi(\tau)d\tau, \qquad (7a)$$

$$\varphi(t) = \int_{-\infty}^{t} c\exp\{-iB(t-\tau)\}\left[(\sigma \cdot \pi)/i\hbar\right]\psi(\tau)d\tau. \qquad (7b)$$

Using equation (7) in (3a), we have

$$i\hbar\frac{\partial \psi}{\partial t} = (V + mc^2)\psi + \left[c^2(\sigma \cdot \pi)/i\hbar\right]\int_{-\infty}^{t} \exp\{-iB(t-\tau)\}(\sigma \cdot \pi)\psi(\tau)d\tau. \qquad (8)$$

In a similar manner, we obtain the complete equation for $\varphi$:

$$i\hbar\frac{\partial \varphi}{\partial t} = (V - mc^2)\varphi + \left[c^2(\sigma \cdot \pi)/i\hbar\right]\int_{-\infty}^{t} \exp\{-iB'(t-\tau)\}(\sigma \cdot \pi)\varphi(\tau)d\tau, \qquad (9)$$

where $B' = \left[(V + mc^2)/\hbar\right]$ and $v(t) = \theta(t)\exp\{-iB't\}$, which allows us to solve for $\psi$:

$$\psi(t) = cv(t) * [(\sigma \cdot \pi)/i\hbar]\varphi(t) = \int_{-\infty}^{t} c\exp\{-iB'(t-\tau)\}[(\sigma \cdot \pi)/i\hbar]\varphi(\tau)d\tau. \qquad (10)$$

Thus, we have decomposed $L^2(\mathbf{R}^3, \mathbf{C}^4)$ as $L^2(\mathbf{R}^3, \mathbf{C}^4) = L^2(\mathbf{R}^3, \mathbf{C}^2) \oplus L^2(\mathbf{R}^3, \mathbf{C}^2)$. One copy of $L^2(\mathbf{R}^3, \mathbf{C}^2)$ contains the particle (positive energy) wave component, while the other copy contains the antiparticle (negative energy) wave component. Which of these copies corresponds to the components $\psi = (\psi_1, \psi_2)^t$ and which to the components $\varphi = (\varphi_1, \varphi_2)^t$ depends, to some extent, on the properties of the scalar potential $V$. However, we will not consider this problem in greater detail in the present paper. An unsettled issue is the definition of the appropriate inner product for the two subspaces, which will account for the quantum constraint that the total probability integral is normalized. We can satisfy this requirement if we set $(\psi, \chi) = \psi_1 \bar{\chi}_1 + \psi_2 \bar{\chi}_2$, and $(\psi, \chi)_A = (A\psi, A\chi)$, $(\varphi, \eta)_{A'} = (A'\varphi, A'\eta)$, where $A\psi = cu(t) * [(\sigma \cdot \boldsymbol{\pi})/i\hbar]\psi(t)$, $A'\varphi = cv(t) * [(\sigma \cdot \boldsymbol{\pi})/i\hbar]\varphi(t)$. Define the particle and antiparticle inner products by

$$\langle \psi, \chi \rangle_p = \int_{\mathbf{R}^3} \{(\psi, \chi) + (\psi, \chi)_A\} d\mathbf{x}, \tag{11a}$$

$$\langle \varphi, \eta \rangle_{ap} = \int_{\mathbf{R}^3} \{(\varphi, \eta) + (\varphi, \eta)_{A'}\} d\mathbf{x}, \tag{12a}$$

so that:

$$\rho_\psi = |\psi|^2 + \left| \int_{-\infty}^{t} c \exp\{-iB(t-\tau)\}[(\sigma \cdot \boldsymbol{\pi})/i\hbar]\psi(\tau)d\tau \right|^2, \tag{11b}$$

$$\rho_\varphi = |\varphi|^2 + \left| \int_{-\infty}^{t} c \exp\{-iB'(t-\tau)\}[(\sigma \cdot \boldsymbol{\pi})/i\hbar]\varphi(\tau)d\tau \right|^2. \tag{12b}$$

It is clear that $\int_{\mathbf{R}^3} \rho_\psi dx = \int_{\mathbf{R}^3} \rho_\varphi dx = 1$, so we now have a complete separation of the particle and antiparticle wave functions.

In the standard representation, the charge conjugation operator is $\mathbf{C}\psi = U_C \bar{\psi}$, with $U_C = i\boldsymbol{\beta}\alpha_2$. A simple computation establishes the following theorem.

**Theorem 1.** *Equations (8) and (9) are mapped into each other under the charge conjugation transformation.*

Equations (8) and (9) offer an interesting alternative to the many attempts to decompose the Dirac equation into particle-antiparticle and/or parity-sensitive pairs. They also offer a different approach to the study of large Z (hydrogen-like) atoms. Although not a part of our direction, one should be able to show that (under physically reasonable conditions) equation (8) is stable in the large Z limit for such atoms.

**IV. Interpretations**

Writing the Dirac equation and the direct separation in two-component matrix form, we have:

$$i\hbar \frac{\partial}{\partial t}\begin{bmatrix} \psi \\ \varphi \end{bmatrix} = \begin{bmatrix} (V+mc^2) & c(\sigma \cdot \boldsymbol{\pi}) \\ c(\sigma \cdot \boldsymbol{\pi}) & (V-mc^2) \end{bmatrix}\begin{bmatrix} \psi \\ \varphi \end{bmatrix},\quad (13)$$

and

$$i\hbar \frac{\partial}{\partial t}\begin{bmatrix} \psi \\ \varphi \end{bmatrix} = \begin{bmatrix} (V+mc^2) \\ +[c^2(\sigma \cdot \boldsymbol{\pi})/i\hbar][u*(\sigma \cdot \boldsymbol{\pi})] & 0 \\ 0 & (V-mc^2) \\ & +[c^2(\sigma \cdot \boldsymbol{\pi})/i\hbar][v*(\sigma \cdot \boldsymbol{\pi})] \end{bmatrix}\begin{bmatrix} \psi \\ \varphi \end{bmatrix}. \quad (14)$$

We call (14) the *analytic diagonalization* of the Dirac equation because the wave function has not changed.

The standard approach to the diagonalization of the Dirac equation (without an external potential $V$) is via the Foldy-Wouthuysen representation. Assuming that **A** does not depend on $t$, the following generalization can be found in deVries[5]:

$$i\hbar \frac{\partial}{\partial t}\begin{bmatrix} \Phi_1 \\ \Phi_2 \end{bmatrix} = \begin{bmatrix} \sqrt{c^2\boldsymbol{\pi}^2 - \frac{e}{c}(\sigma \cdot \mathbf{B}) + m^2c^4} & 0 \\ 0 & -\sqrt{c^2\boldsymbol{\pi}^2 - \frac{e}{c}(\sigma \cdot \mathbf{B}) + m^2c^4} \end{bmatrix}\begin{bmatrix} \Phi_1 \\ \Phi_2 \end{bmatrix}. \quad (15)$$

In this case, $[\Phi_1 \ \Phi_2]^t = U_{FW}[\psi \ \varphi]^t$, and $\mathbf{H}_s = U_{FW}\mathbf{H}_D U_{FW}^{-1}$ (see Thaller[3]). Equation (15) will be studied in detail elsewhere[21]. However, it is known (see Gill[10]) that when **A** is zero, $\mathbf{H}_s = \boldsymbol{\beta}\sqrt{c^2\mathbf{p}^2 + m^2c^4}$ has the following analytic representation:

$$\mathbf{H}_s f(\mathbf{x}) =$$
$$-\frac{\mu^2\hbar^2 c\boldsymbol{\beta}}{\pi^2}\int_{\mathbf{R}^3}\left\{\left[\frac{\mathbf{K}_0[\mu\|\mathbf{x}-\mathbf{y}\|]}{\|\mathbf{x}-\mathbf{y}\|} + \frac{2\mathbf{K}_1[\mu\|\mathbf{x}-\mathbf{y}\|]}{\mu\|\mathbf{x}-\mathbf{y}\|^2}\right]\left[\frac{1}{\|\mathbf{x}-\mathbf{y}\|} - 2\pi\delta(\mathbf{x}-\mathbf{y})\right]\right\}f(\mathbf{y})d\mathbf{y}. \quad (16)$$

Here, the $\mathbf{K}_n$ are modified Bessel functions of the third kind and $\mu = mc/\hbar$. Equation (16) is of independent interest, since it is the first example of a physically relevant operator which has a representation as the confinement of a composite of three singularities, two negative and one (hard core) positive, within a Compton wavelength such that, at the point of singularity, they cancel each other providing a finite result. The second paper in this series is devoted to the square root operator, where this result discussed in detail.

We can now interpret the zitterbewegung, and the result that a velocity measurement (of a Dirac particle) at any instant in time is $\pm c$, as reflections of the fact that the Dirac equation makes a spatially extended particle appear as a point in the present by forcing it to oscillate between the past and future at speed $c$.

From equation (14), we conclude that the coupling of the particle and antiparticle wave functions in the first-order form of the Dirac equation hides the second order nonlocal time nature of the equation. From (16), we see explicitly that (15) is nonlocal in space. Thus, the implicit time nonlocality of the Dirac equation is mapped into the explicit spatial nonlocality of the square-root equation by the Foldy-Wouthuysen transformation. These observations imply that the Dirac Hamiltonian $\mathbf{H}_D$ and the square-root Hamiltonian, $\mathbf{H}_s = U_{FW} \mathbf{H}_D U_{FW}^{-1}$, are mathematically, but not (what we mean by) physically equivalent. Furthermore, it appears that the only way we can partially justify using the square-root equation to interpret the Dirac equation is that they both square to give the Klein-Gordon equation. We conclude that they can only be viewed as physically equivalent outside a Compton wavelength where they both appear as point particles.

## V. The Hydrogen Atom

In this section, we reconsider the standard analysis of the Dirac equation for the hydrogen atom problem from an exact point of view. We assume that $\mathbf{A} = (\boldsymbol{\mu}_I \times \mathbf{r})/r^3$, $V = -\hbar c \gamma / r$, and $\gamma = e^2/\hbar c$. Rewrite (3a) and (3b) in eigenvalue form:

$$(E - V - mc^2)\psi = c(\boldsymbol{\sigma} \cdot \boldsymbol{\pi})\varphi, \tag{17a}$$

$$(E - V + mc^2)\varphi = c(\boldsymbol{\sigma} \cdot \boldsymbol{\pi})\psi. \tag{17b}$$

Eliminating $\varphi$ in terms of $\psi$ and vice versa, we obtain the following equations:

$$(E - V - mc^2)\psi = \frac{c^2(\boldsymbol{\sigma} \cdot \mathbf{p}V)(\boldsymbol{\sigma} \cdot \boldsymbol{\pi})}{(E - V + mc^2)^2}\psi + \frac{c^2(\boldsymbol{\sigma} \cdot \boldsymbol{\pi})(\boldsymbol{\sigma} \cdot \boldsymbol{\pi})}{(E - V + mc^2)}\psi, \tag{18a}$$

$$(E - V + mc^2)\varphi = \frac{c^2(\boldsymbol{\sigma} \cdot \mathbf{p}V)(\boldsymbol{\sigma} \cdot \boldsymbol{\pi})}{(E - V - mc^2)^2}\varphi + \frac{c^2(\boldsymbol{\sigma} \cdot \boldsymbol{\pi})(\boldsymbol{\sigma} \cdot \boldsymbol{\pi})}{(E - V - mc^2)}\varphi. \tag{18b}$$

(We also get (18) from equations (8) and (9) via straightforward integration, using the Riemann-Lebesgue Lemma.) We call (18a) and (18b) the Slater equations since they were first used by one of his students as early as 1940[12], and appeared in his book[13], first published in 1960 (see Appendix 29). (It's surprising that Slater's work is not well known.) For obvious reasons, we concentrate on (18a). First note that, if we drop the middle term and replace $(E - V + mc^2)$ by $2mc^2$, we get the Pauli approximation to the Dirac equation:

$$(E - V - mc^2)\psi = -\frac{e\hbar}{2mc}(\boldsymbol{\sigma} \cdot \mathbf{B})\psi + \frac{\boldsymbol{\pi}^2}{2m}\psi. \tag{19}$$

As noted earlier, the Pauli equation was used to extract the hyperfine splitting portion of the hydrogen spectrum to support the predictions of QED. It follows that the conditions that justify the Pauli approximation and the dropping of the middle term of (18a) are both of importance for the foundations.

There are a number of other equations and/or approximations that have been given the name and/or used in lieu of the Pauli equation (see for example, Greiner[14], Mizushima[15] or Bethe and Salpeter[16]). We do not consider these equations since, although they are related to the Dirac equation, they do not give additional information and it is not obvious that they have any more mathematical or physical justification when applied to the s-states of hydrogen.

Recall that there is a finite probability of finding the electron at the origin in s-states, but the required condition for the validity of (19) is $(E - V + mc^2) \ll 2mc^2$. Thus, this condition is not satisfied for s-state calculations. It follows that use of the Pauli equation to compute the hyperfine splitting of s-states is not convincing. On the other hand, the condition is easily seen to be satisfied for all other states. A more reasonable approximation is to use $|mc^2 - E| \ll mc^2$ to replace $(E - V + mc^2)$ by $2mc^2(1 + r_0/r)$, where $r_0 = e^2/(E + mc^2) \cong e^2/2mc^2$. The above condition is always satisfied ($13 ev$ compared to $0.5 Mev$). This approach also has the additional advantage of removing the nonlinear eigenvalue problem posed by (18a) without substantially affecting the final result. In this case we have

$$(E - V - mc^2)\psi = \frac{(\sigma \cdot \mathbf{p}V)(\sigma \cdot \boldsymbol{\pi})}{4m^2c^2(1 + r_0/r)^2}\psi + \frac{(\sigma \cdot \boldsymbol{\pi})(\sigma \cdot \boldsymbol{\pi})}{2m(1 + r_0/r)}\psi. \tag{20}$$

Using standard computations, we get (see Slater[13], $\hbar \mathbf{L} = \mathbf{r} \times \mathbf{p}$ is the angular momentum, and $\hbar \mathbf{S}$ is the spin, $\mathbf{S} = \sigma/2$)

$$(\sigma \cdot \boldsymbol{\pi})(\sigma \cdot \boldsymbol{\pi}) = \boldsymbol{\pi}^2 - \frac{2e\hbar}{c}\left\{\frac{8\pi}{3}(\mathbf{S} \cdot \boldsymbol{\mu}_I)\delta(\mathbf{r}) + \left[\frac{3(\mathbf{S} \cdot \mathbf{r})(\boldsymbol{\mu}_I \cdot \mathbf{r})}{r^5} - \frac{(\mathbf{S} \cdot \boldsymbol{\mu}_I)}{r^3}\right]\right\}, \tag{21a}$$

$$(\sigma \cdot \mathbf{p}V)(\sigma \cdot \boldsymbol{\pi}) = \frac{2e^2\hbar}{r^2}\left\{\hbar\left[\frac{(\mathbf{S} \cdot \mathbf{L})}{r} - \frac{d}{dr}\right] + \frac{e}{c}\left[\frac{(\mathbf{S} \cdot \boldsymbol{\mu}_I)}{r^2} - \frac{(\mathbf{S} \cdot \mathbf{r})(\boldsymbol{\mu}_I \cdot \mathbf{r})}{r^4}\right]\right\}. \tag{21b}$$

Putting these expressions in (20), we have:

$$(E - V - mc^2)\psi = \frac{r_0 \hbar}{m(1 + r_0/r)^2 r^2}\left\{\hbar\left[\frac{(\mathbf{S} \cdot \mathbf{L})}{r} - \frac{d}{dr}\right] + \frac{e}{c}\left[\frac{(\mathbf{S} \cdot \boldsymbol{\mu}_I)}{r^2} - \frac{(\mathbf{S} \cdot \mathbf{r})(\boldsymbol{\mu}_I \cdot \mathbf{r})}{r^4}\right]\right\}\psi$$
$$- \frac{e\hbar}{mc(1 + r_0/r)}\left\{\frac{8\pi}{3}(\mathbf{S} \cdot \boldsymbol{\mu}_I)\delta(\mathbf{r}) + \left[\frac{3(\mathbf{S} \cdot \mathbf{r})(\boldsymbol{\mu}_I \cdot \mathbf{r})}{r^5} - \frac{(\mathbf{S} \cdot \boldsymbol{\mu}_I)}{r^3}\right]\right\}\psi + \frac{\boldsymbol{\pi}^2}{2m(1 + r_0/r)}\psi \tag{22}$$

When $\boldsymbol{\mu}_I = 0$, (22) becomes

$$(E - V - mc^2)\psi = \frac{r_0 \hbar^2}{m(1 + r_0/r)^2 r^2}\left[\frac{(\mathbf{S}\cdot\mathbf{L})}{r} - \frac{d}{dr}\right]\psi + \frac{\mathbf{p}^2}{2m(1 + r_0/r)}\psi. \quad (23)$$

Equation (23) has (using $r_0 = e^2/(E + mc^2)$) the same eigenvalues as the unperturbed Dirac equation, so that our interest centers on the following terms (op means operator)

$$-\frac{e\hbar}{mc(1 + r_0/r)}\left\{\frac{8\pi}{3}(\mathbf{S}\cdot\boldsymbol{\mu}_I)\delta(\mathbf{r}) + \left[\frac{3(\mathbf{S}\cdot\mathbf{r})(\boldsymbol{\mu}_I\cdot\mathbf{r})}{r^5} - \frac{(\mathbf{S}\cdot\boldsymbol{\mu}_I)}{r^3}\right]_{op}\right\}, \quad (24a)$$

$$\frac{er_0\hbar}{mc(1 + r_0/r)^2 r^2}\left[\frac{(\mathbf{S}\cdot\boldsymbol{\mu}_I)}{r^2} - \frac{(\mathbf{S}\cdot\mathbf{r})(\boldsymbol{\mu}_I\cdot\mathbf{r})}{r^4}\right]_{op}. \quad (24b)$$

The delta term in equation (24a), except for the additional factor $(1 + r_0/r)^{-1}$, would normally be used to compute the hyperfine splitting of s-states in the Pauli approximation. It is easy to see that, with this additional factor, the same calculation would give a value of zero for the splitting. In all other states this factor is small ($1 \gg r_0/r$) and may be dropped.

Slater[13] used equation (24b) to compute the s-state (hyperfine) splitting and obtained the correct result. Since this term is (part of the) focus of our investigation, we repeat some of Slater's calculations. In the s-state the total angular momentum $\mathbf{J}$ is equal to $\mathbf{S}$. Hence, following standard procedures, we replace $\left[(\mathbf{S}\cdot\boldsymbol{\mu}_I/r^2) - ((\mathbf{S}\cdot\mathbf{r})(\boldsymbol{\mu}_I\cdot\mathbf{r})/r^4)\right]_{op}$ by $(\mathbf{S}\cdot\boldsymbol{\mu}_I/\mathbf{S}^2)\left[(\mathbf{S}^2/r^2) - ((\mathbf{S}\cdot\mathbf{r})^2/r^4)\right]_{op}$. It is easy to see that, as far as operator averages are concerned, $\left[(\mathbf{S}\cdot\mathbf{r})^2\right]_{op} = \frac{1}{4}(r)^2_{op}$ and $\left[\mathbf{S}^2\right]_{op} = \frac{3}{4}$. The term of interest becomes

$$\frac{2er_0\hbar}{3mc(1 + r_0/r)^2 r^4}(\mathbf{S}\cdot\boldsymbol{\mu}_I)_{op}. \quad (25a)$$

The important issue is the computation of the s-state expected value of

$$\frac{r_0\lambda}{(1 + r_0/r)^2 r^4}, \quad (25b)$$

where $\lambda = 2e\hbar/3mc\left\langle(\mathbf{S}\cdot\boldsymbol{\mu}_I)_{op}\right\rangle_{ave}$. Slater[13] assumed the nonrelativistic radial wave function for s-states. (For the $2s_{1/2}$ state, $R(r) = \frac{1}{\sqrt{2}}\eta^{3/2}(1 - \frac{1}{2}r\eta)\exp(-\frac{1}{2}r\eta)$ and $\eta = 1/r_B$, where $r_B = 0.529178 \times 10^{-10} m$ is the Bohr radius.) Using the normalization $\int_0^\infty r^2 R(r)^2 dr = 1$, this led him to the computation of

$$\lambda \int_0^\infty \frac{r_0 R(r)^2 r^2}{(1+r_0/r)^2 r^4} dr. \tag{26a}$$

Setting $\rho = \eta r$ and $\rho_0 = \eta r_0$, we have

$$\tfrac{1}{2}\eta^3 \rho_0 \lambda \int_0^\infty \frac{(1-\tfrac{1}{2}\rho)^2 \exp(-\rho) d\rho}{(1+\rho_0/\rho)^2 \rho^2}. \tag{26b}$$

By a change of variables ($u = \rho + \rho_0$) and integration by parts, it is easy to see that $\rho_0$ is a cutoff and that the dominant contribution to the expression (26b) is $\tfrac{1}{2}\lambda\eta^3$ as $\rho_0 \to 0^+$. (Note that the integral is divergent but the factor $\rho_0$ in front makes the product finite.) We get the same result for all s-states, while it is not hard to show that equation (24b) is (almost) zero for all other states.

It would appear that the correct approach for s-state (hyperfine) splitting gives the same results as those obtained from the Pauli equation. Furthermore, equation (24b) introduces a natural cutoff, which removes the conceptual difficulty of a point magnetic dipole interaction as implied by use of the delta term in the Pauli equation. In addition, it is not hard to show that Slater's approach goes through, giving the same result, if we use (the correct) Dirac solution for the first-order calculation.

However, to provide input for the precise results of QED, we must first correct the normalization condition to

$$\int \left[|\psi|^2 + |(\sigma \cdot \mathbf{p})\psi|^2 / 4m^2c^2(1+r_0/r)^2\right] d\mathbf{x} = 1. \tag{27}$$

Clearly, we expect that the additional term will give a very small correction. However, it is not clear how small. For example, if it changes the hyperfine splitting values in the eight or ninth decimal place (in GHz), it may well be an important correction. (For example, the measured value of the $2s_{1/2}$ state hyperfine splitting in hydrogen is 0.177566850(10) Ghz, see Mizushima[15].)

We now approach the more difficult issue facing attempts to completely understand the Dirac problem for full coupling, namely, the $\mathbf{A}^2$ term:

$$\frac{e^2 \mathbf{A}^2}{2mc^2(1+r_0/r)} = \frac{e^2 \mu_I^2 \sin^2\theta}{2mc^2(1+r_0/r)r^4}. \tag{28}$$

In most treatments of the Dirac hydrogen atom problem, this term (with $r_0 = 0$) is either ignored or assumed to be small. Power counting shows that it cannot be ignored without investigation. It is easy to show that this term will be small in all except s-states.

The first observation is that this term appears to be more singular than the Coulomb potential, so that perturbation analysis may not be appropriate. However, this is not completely clear since the $\sin^2\theta$ term vanishes on the spin axis and could strongly modify the singular nature of this term.

If we take an engineering approach and assume that we can treat the $\mathbf{A}^2$ term as a perturbation, then for the 2s- state the expected value is

$$\int_0^\infty \frac{r_0 \mu_I^2 R(r)^2 r^2}{(1+r_0/r)r^4} dr \int_0^\pi \sin^2\theta(\sin\theta)d\theta = \tfrac{1}{3}\eta^3 \rho_0 \mu_I^2 \int_0^\infty \frac{(1-\tfrac{1}{2}\rho)^2 \exp(-\rho)d\rho}{(1+\rho_0/\rho)\rho^2}. \tag{29}$$

In atomic units, $\eta = 1$, $\mu_0 = (1/2)\gamma$, $r_0 = \rho_0 = (1/2)\gamma^2$, $g_N^2 = 30.9136$, $\langle \mathbf{I}_{op}^2 \rangle_{ave} = (3/4)$ and $\mu_I^2 = (1/1836)^2 g_N^2 \mu_0^2 \mathbf{I}_{op}^2$, so we can write (29) as

$$\tfrac{1}{3}\mu_I^2 \int_0^\infty \left[\frac{1}{\rho} + \frac{\rho_0}{4} - \frac{(1+\rho_0+\tfrac{1}{4}\rho_0^2)}{(\rho_0+\rho)}\right]\exp(-\rho)d\rho. \tag{30}$$

Using a table of integrals (see Gradshteyn and Ryzhik[17], pages 925 and 927) and the cutoff prescription of Bethe[16] (page 110), we have $\int_\varepsilon^\infty (1/\rho)\exp(-\rho)d\rho = -Ei(-\varepsilon)$, and $-\int_0^\infty 1/(\rho_0+\rho)\exp(-\rho)d\rho = e^{\rho_0} Ei(-\rho_0)$, where $Ei(-\varepsilon) = C + \ln\varepsilon + \sum_{k=1}^\infty (-1)^k [\varepsilon^k/k(k!)]$ and $C$ is Euler's constant. Using these results in (30), we get

$$\tfrac{1}{3}\mu_I^2 \left\{ \lim_{\varepsilon\to 0} \int_\varepsilon^\infty \frac{1}{\rho}\exp(-\rho)d\rho + \int_0^\infty \left[\frac{\rho_0}{4} - \frac{(1+\rho_0+\tfrac{1}{4}\rho_0^2)}{(\rho_0+\rho)}\right]\exp(-\rho)d\rho \right\}$$

$$= \tfrac{1}{3}\mu_I^2 \left[ -\lim_{\varepsilon\to 0} Ei(-\varepsilon) + \frac{\rho_0}{4} + e^{\rho_0}(1+\rho_0+\tfrac{1}{4}\rho_0^2)Ei(-\rho_0) \right]. \tag{31}$$

It is clear that $-Ei(-\varepsilon)$ will diverge like $-\ln\varepsilon$ as $\varepsilon \to 0$. If we fix $\varepsilon$ at $\rho_0$, and note that $e^{\rho_0} \cong 1+\rho_0$, then

$$\tfrac{1}{3}\mu_I^2 \left[ -Ei(-\rho_0) + \frac{\rho_0}{4} + e^{\rho_0}(1+\rho_0+\tfrac{1}{4}\rho_0^2)Ei(-\rho_0) \right]$$

$$\cong \tfrac{1}{3}\mu_I^2 \left\{ \frac{\rho_0}{4} + (2\rho_0+\tfrac{5}{4}\rho_0^2)\left[C + \ln\rho_0 + \sum_{k=1}^\infty \frac{(-1)^k \rho_0^k}{k(k!)}\right]\right\} \Rightarrow$$

$$\tfrac{1}{3}\rho_0\mu_I^2 \int_0^\infty \frac{(1-\tfrac{1}{2}\rho)^2 \exp(-\rho)d\rho}{(1+\rho_0/\rho)\rho^2} \cong \tfrac{1}{16}\gamma^2(\tfrac{1}{1836})^2 g_N^2 \left\{\tfrac{1}{8}\gamma^2 + (\gamma^2+\tfrac{5}{16}\gamma^4)\left[C+\ln\tfrac{1}{2}\gamma^2 - \tfrac{1}{2}\gamma^2\right]\right\}. \tag{32}$$

If we note that $(1/1836)^2 \cong (1/13)^2 \gamma^2$, then this last term is of order $(>) \gamma^7$. Thus, if there is (mathematical) support for the calculation procedures, the $\mathbf{A}^2$ term does not make a significant contribution. In the next section, we discuss an exact separation of variables for full coupling, Coulomb plus magnetic dipole vector potentials, which will allow us to partially answer this question.

## VI. Separation of Variables

In this section, we show that an exact separation of variables is possible via a new approach based on earlier work of Harish-Chandra[18] and Villalba[19]. To begin, we reconsider equation (3) in eigenvalue form with $\mathbf{A} = (\boldsymbol{\mu}_I \times \mathbf{r})/r^3$, $V = -\hbar c \gamma / r$ and $\gamma = Ze^2/\hbar c$. Following Dirac[20], set $\mathbf{I}_4 = diag(1,1,1,1)$, $\alpha_i = \rho_1 \Sigma_i$, $\rho_3 = \boldsymbol{\beta}$, where

$$\Sigma_i = \begin{bmatrix} \sigma_i & 0 \\ 0 & \sigma_i \end{bmatrix}, \quad \rho_1 = \begin{bmatrix} 0 & \mathbf{I}_2 \\ \mathbf{I}_2 & 0 \end{bmatrix}, \quad \rho_2 = \begin{bmatrix} 0 & -i\mathbf{I}_2 \\ i\mathbf{I}_2 & 0 \end{bmatrix}. \tag{33}$$

It follows that $\rho_i \Sigma_j = \Sigma_j \rho_i$ $(i,j = 1,2,3)$, and equation (3) can be written in the form:

$$\mathbf{H}_D \Psi = E \Psi = \left[ c \rho_1 \boldsymbol{\Sigma} \cdot (\mathbf{p} - \tfrac{e}{c}\mathbf{A}) + mc^2 \rho_3 + V\mathbf{I}_4 \right] \Psi. \tag{34}$$

Following Harish-Chandra[18], we represent the components of $\boldsymbol{\Sigma}$ in spherical coordinates:

$$\begin{aligned} \Sigma_r &= (\Sigma_1 \cos\varphi + \Sigma_2 \sin\varphi)\sin\theta + \Sigma_3 \cos\theta \\ \Sigma_\theta &= r(\Sigma_1 \cos\varphi + \Sigma_2 \sin\varphi)\cos\theta - r\Sigma_3 \sin\theta \\ \Sigma_\varphi &= r(\Sigma_2 \cos\varphi - \Sigma_1 \sin\varphi)\sin\theta \end{aligned} \tag{35}$$

so that

$$\boldsymbol{\Sigma} \cdot \mathbf{p} = \Sigma_r p_r + \frac{\Sigma_\theta p_\theta}{r} + \frac{\Sigma_\varphi p_\varphi}{r \sin\theta}. \tag{36}$$

With the proton's magnetic moment lying along the z-axis, the vector potential (in spherical coordinates) takes the form $A_r = A_\theta = 0$ and $A_\varphi = (|\boldsymbol{\mu}_I|/r^2)\sin\theta$, so that:

$$\boldsymbol{\Sigma} \cdot \mathbf{A} = \Sigma_\varphi A_\varphi = \frac{\Sigma_\varphi |\boldsymbol{\mu}_I|}{r^2} \sin\theta. \tag{37}$$

It is well known that the orbital angular momentum $L_z = -i\partial/\partial\varphi$ about the z-axis is not a conserved quantity (taking $\hbar = 1$ for now). However, it is easy to show that the total angular momentum about the z-axis, $L_z \mathbf{I}_4 + \tfrac{1}{2}\Sigma_3$, commutes with $\mathbf{H}_D$:

$$[\mathbf{H}_D, L_z\mathbf{I}_4 + \tfrac{1}{2}\Sigma_3] = \mathbf{0}, \tag{38}$$

so that it is conserved. Therefore, we can choose solutions of (34) in the form:

$$\Psi(r,\theta,\varphi) = \exp[i(m - \tfrac{1}{2}\Sigma_3)\varphi]\Psi^1(r,\theta), \tag{39}$$

where $\Psi^1(r,\theta)$ does not depend on $\varphi$ and $m$ is a half odd integer with both positive and negative values allowed.

Now using:

$$\begin{aligned}
\Sigma_r &= \exp\{-\tfrac{i}{2}\Sigma_3\varphi\}\exp\{-\tfrac{i}{2}\Sigma_2\theta\}\Sigma_3\exp\{\tfrac{i}{2}\Sigma_2\theta\}\exp\{\tfrac{i}{2}\Sigma_3\varphi\} \\
\Sigma_\theta &= r\exp\{-\tfrac{i}{2}\Sigma_3\varphi\}\exp\{-\tfrac{i}{2}\Sigma_2\theta\}\Sigma_1\exp\{\tfrac{i}{2}\Sigma_2\theta\}\exp\{\tfrac{i}{2}\Sigma_3\varphi\} \\
\Sigma_\varphi &= r\sin\theta\exp\{-\tfrac{i}{2}\Sigma_3\varphi\}\exp\{-\tfrac{i}{2}\Sigma_2\theta\}\Sigma_2\exp\{\tfrac{i}{2}\Sigma_2\theta\}\exp\{\tfrac{i}{2}\Sigma_3\varphi\}
\end{aligned} \tag{40}$$

and the easily–proved relations:

$$\exp\{\tfrac{i}{2}\Sigma_2\theta\}\frac{\partial}{\partial\theta} = \left(\frac{\partial}{\partial\theta} - \frac{i}{2}\Sigma_2\right)\exp\{\tfrac{i}{2}\Sigma_2\theta\} \tag{41a}$$

and

$$\exp\{\tfrac{i}{2}\Sigma_2\theta\}\exp\{\tfrac{i}{2}\Sigma_3\varphi\}\frac{\partial}{\partial\varphi}$$
$$= \left\{\frac{\partial}{\partial\varphi} - \frac{i}{2}(\Sigma_3\cos\theta - \Sigma_1\sin\theta)\right\}\exp\{\tfrac{i}{2}\Sigma_2\theta\}\exp\{\tfrac{i}{2}\Sigma_3\varphi\}, \tag{41b}$$

equation (34) takes the form:

$$\left[-ic\rho_1\left\{\Sigma_3\left(\frac{\partial}{\partial r} + \frac{1}{r}\right) + \frac{\Sigma_1}{r}\left[\left(\frac{\partial}{\partial\theta} + \frac{1}{2}\cot\theta\right)\mathbf{I}_4 - \Sigma_3\left(m\csc\theta - \frac{e}{c}|\mathbf{\mu}_I|\sin^2\theta\right)\right]\right\} \right.$$
$$\left. + mc^2\rho_3 + (V-E)\mathbf{I}_4\right]\Psi_0 = \mathbf{0}, \tag{42}$$

where $\Psi_0(r,\theta,\varphi) \equiv \exp\{\tfrac{i}{2}\Sigma_2\theta + im\varphi\mathbf{I}_4\}\Psi^1(r,\theta)$.

In order to complete our separation, we now follow the method due to Villalba[19]. Setting $\Psi_0 = \Omega\Phi$, where $\Omega$ is (an operator) to be determined so that

$$K_1\Phi = i\left[-ic\rho_1 r\Sigma_3\left(\frac{\partial}{\partial r} + \frac{1}{r}\right) + r\left(mc^2\rho_3 + (V-E)\mathbf{I}_4\right)\right]\Psi_0 = -\lambda\Phi \tag{43a}$$

and

$$K_2\Phi = i(-ic\rho_1)\Sigma_1\left[\left(\frac{\partial}{\partial\theta}+\frac{1}{2}\cot\theta\right)\mathbf{I}_4 - \Sigma_3\left(m\csc\theta-\frac{e}{c}|\boldsymbol{\mu}_\mathbf{I}|\sin^2\theta\right)\right]\Psi_0 = \lambda\Phi. \quad (43b)$$

Then $(K_1 + K_2)\Phi = \mathbf{0}$. Using our definition of $\Psi_0$, we have

$$K_1 = c\rho_1 r\left(\frac{\partial}{\partial r}+\frac{1}{r}\right)\Sigma_3\Omega + irmc^2\rho_3\Omega + ir(V-E)\Omega \quad (44a)$$

and

$$K_2 = c\left[\left(\frac{\partial}{\partial\theta}+\frac{1}{2}\cot\theta\right)\rho_1\Sigma_1\Omega + i\Sigma_2\Omega\left(m\csc\theta - \frac{e}{c}|\mathbf{m}_\mathbf{I}|\sin^2\theta\right)\right]. \quad (44b)$$

We obtain a complete separation provided that $[K_1, K_2] = \mathbf{0}$. It is easy to see that this requirement imposes the following conditions (recall that $\alpha_i = \rho_1\Sigma_i$)

1. $[\alpha_3\Omega, \alpha_1\Omega] = \mathbf{0}$,   2. $[\alpha_3\Omega, \alpha_2\Omega] = \mathbf{0}$,   3. $[\rho_3\Omega, \alpha_1\Omega] = \mathbf{0}$,   (45a)

1. $[\rho_3\Omega, \alpha_2\Omega] = \mathbf{0}$,   2. $[\Omega, \alpha_1\Omega] = \mathbf{0}$,   3. $[\Omega, \alpha_2\Omega] = \mathbf{0}$.   (45b)

If $\Omega^{-1}$ exists, then from (45b)-2 we obtain $[\Omega, \alpha_1] = \mathbf{0}$ and from (45b)-3, $[\Omega, \alpha_2] = \mathbf{0}$. From (45a)-1,

$$\begin{aligned}[\alpha_3\Omega, \alpha_1\Omega] &= [\alpha_3, \alpha_1]\Omega^2 + \alpha_1[\alpha_3, \Omega]\Omega \\ &= 2i\Sigma_2\Omega^2 + \alpha_1[\alpha_3, \Omega]\Omega.\end{aligned} \quad (46)$$

Then, since $\alpha_i^{-1} = \alpha_i$ $(i = 1,2,3)$,

$$\begin{aligned}[\alpha_3, \Omega] &= -2i\alpha_1\Sigma_2\Omega = -2i\rho_1\Sigma_1\Sigma_2\Omega \\ &= -2i\rho_1(i\Sigma_3)\Omega = 2\alpha_3\Omega \Rightarrow \{\alpha_3, \Omega\} = \mathbf{0}\end{aligned}$$

(i.e., $\alpha_3$ and $\Omega$ anticommute). Similar procedures yield the same conditions from (45a)-2 and (45b)-1. Now write

$$\Omega = \begin{bmatrix} A & B \\ C & D \end{bmatrix},$$

where $A, B, C$ and $D$ are $2\times 2$ matrices. Expanding these matrices as series of the form

$$A = a_0\mathbf{I}_2 + \sum_{i=1}^{3} a_i\sigma_i,$$

with similar expressions for $B, C$ and $D$, and using $[\Omega, \alpha_1] = [\Omega, \alpha_2] = \{\Omega, \alpha_3\} = 0$ shows that

$$\Omega = b \begin{bmatrix} 0 & \sigma_3 \\ -\sigma_3 & 0 \end{bmatrix}, \quad b = const. \tag{47}$$

Setting $b$ equal to 1, this completes the separation of variables for system (42).

We now consider the eigenvalue equations:

$$K_1 \Phi = \left\{ c\rho_1 r \left( \frac{\partial}{\partial r} + \frac{1}{r} \right) \Sigma_3 \Omega + ir\, mc^2 \rho_3 \Omega + ir(V - E)\Omega \right\} \Phi = -\lambda \Phi, \tag{48}$$

$$K_2 \Phi = c \left[ \left( \frac{\partial}{\partial \theta} + \frac{1}{2} \cot\theta \right) \rho_1 \Sigma_1 \Omega + i\rho_1 \Sigma_2 \Omega \left( m\csc\theta - \frac{e}{c} |m_I| \sin^2\theta \right) \right] \Phi = \lambda \Phi, \tag{49}$$

for the radial and angular equations respectively.

**Solution of the Radial Equations**

Putting $\Phi = [\Phi_1, \Phi_2]^t$, with $\Phi_1$ and $\Phi_2$ 2-spinors, we obtain from equation (48) (reinstating the relevant factors of $\hbar$):

$$-r \left( \frac{\partial}{\partial r} + \frac{1}{r} \right) \Phi_1 + ir \left( \frac{mc^2 - E}{\hbar c} - \frac{\gamma}{r} \right) \sigma_3 \Phi_2 = -\frac{\lambda}{\hbar c} \Phi_1, \tag{50a}$$

$$r \left( \frac{\partial}{\partial r} + \frac{1}{r} \right) \Phi_2 + ir \left( \frac{mc^2 + E}{\hbar c} + \frac{\gamma}{r} \right) \sigma_3 \Phi_1 = -\frac{\lambda}{\hbar c} \Phi_2. \tag{50b}$$

Now put $\Phi_1 = \varsigma a(r)$, $\Phi_2 = \eta b(r)$, where $\varsigma, \eta$ are two-spinors and $a(r), b(r)$ are functions. Substitution into equations (50a) and (50b) gives:

$$-r \left( \frac{\partial}{\partial r} + \frac{1}{r} \right) \varsigma a(r) + ir \left( \frac{mc^2 - E}{\hbar c} - \frac{\gamma}{r} \right) (\sigma_3 \eta) b(r) = -\frac{\lambda}{\hbar c} \varsigma a(r),$$

$$r \left( \frac{\partial}{\partial r} + \frac{1}{r} \right) \eta b(r) + ir \left( \frac{mc^2 + E}{\hbar c} + \frac{\gamma}{r} \right) (\sigma_3 \varsigma) a(r) = -\frac{\lambda}{\hbar c} \eta b(r).$$

If we choose $\eta = [\eta_1, \eta_2]^t$ and $\varsigma = [\eta_1, -\eta_2]^t$, we can eliminate the spinors from the above equations to get ($\sigma_3 \eta = \varsigma$, $\sigma_3 \varsigma = \eta$):

$$-r\left(\frac{\partial}{\partial r} + \frac{1}{r}\right)a(r) + ir\left(\frac{mc^2 - E}{\hbar c} - \frac{\gamma}{r}\right)b(r) = -\frac{\lambda}{\hbar c}a(r), \qquad (51a)$$

$$r\left(\frac{\partial}{\partial r} + \frac{1}{r}\right)b(r) + ir\left(\frac{mc^2 + E}{\hbar c} + \frac{\gamma}{r}\right)a(r) = -\frac{\lambda}{\hbar c}b(r). \qquad (51b)$$

We can now apply standard methods, putting $a(r) = u(r)/r$, $b(r) = v(r)/r$ in (51a) and (51b) yields the equations:

$$-\frac{du}{dr} + \frac{\lambda}{\hbar c}\frac{u}{r} + i\left(\frac{mc^2 - E}{\hbar c} - \frac{\gamma}{r}\right)v = 0, \qquad (52a)$$

$$\frac{dv}{dr} + \frac{\lambda}{\hbar c}\frac{v}{r} + i\left(\frac{mc^2 + E}{\hbar c} + \frac{\gamma}{r}\right)u = 0. \qquad (52b)$$

These equations are very close to those for the Dirac-Coulomb problem (except for the factor of *i*). As in that problem, make the change of independent variable $y = 2\varepsilon r$, with $\varepsilon = \sqrt{m^2 c^4 - E^2}/\hbar c$ and change dependent variables via: (we follow Greiner[14], p. 179)

$$u = e^{-y/2}\sqrt{mc^2 - E}\,(\phi_1 - \phi_2),$$

$$v = e^{-y/2}\sqrt{mc^2 + E}\,(\phi_1 + \phi_2),$$

from which the following equations for $\phi_1$ and $\phi_2$ are obtained:

$$\frac{d\phi_1}{dy} = \left(\frac{1}{2} - i\frac{\gamma}{\hbar c \varepsilon y}mc^2\right)\phi_1 + \left(\frac{-\lambda}{\hbar c y} + \frac{i}{2\hbar c} - i\frac{\gamma}{\hbar c \varepsilon y}E\right)\phi_2, \qquad (53a)$$

$$\frac{d\phi_2}{dy} = -\left(\frac{\lambda}{\hbar c y} + \frac{i}{2\hbar c} - i\frac{\gamma}{\hbar c \varepsilon y}E\right)\phi_1 + \left(\frac{1}{2} + i\frac{\gamma}{\hbar c \varepsilon y}mc^2\right)\phi_2. \qquad (53b)$$

We now look for series solutions of these equations in the form:

$$\phi_1 = \sum_{q=0}^{\infty}\alpha_q y^{q+\delta}, \quad \phi_2 = \sum_{q=0}^{\infty}\beta_q y^{q+\delta},$$

where $\delta$ is to be determined. Substitution into equations (53) gives the following relations:

$$\sum_{q=0}^{\infty} \beta_q \left( q + \delta - i\frac{\gamma mc^2}{\hbar c \varepsilon} \right) y^{q+\delta-1}$$
$$= -\frac{i}{2} \sum_{q=1}^{\infty} \alpha_{q-1} y^{q+\delta-1} - \left( \frac{\lambda}{\hbar c} - i\frac{\gamma E}{\hbar c \varepsilon} \right) \sum_{q=0}^{\infty} \alpha_q y^{q+\delta-1} + \frac{1}{2} \sum_{q=0}^{\infty} \beta_{q-1} y^{q+\delta-1}, \quad (54a)$$

$$\sum_{q=0}^{\infty} \alpha_q \left( q + \delta + i\frac{\gamma mc^2}{\hbar c \varepsilon} \right) y^{q+\delta-1}$$
$$= \frac{1}{2} \sum_{q=1}^{\infty} \alpha_{q-1} y^{q+\delta-1} + \frac{i}{2\hbar c} \sum_{q=1}^{\infty} \beta_{q-1} y^{q+\delta-1} - \left( \frac{\lambda}{\hbar c} + i\frac{\gamma E}{\hbar c \varepsilon} \right) \sum_{q=0}^{\infty} \beta_q y^{q+\delta-1}. \quad (54b)$$

For $q = 0$, equations (54a) and (54b) give

$$\left( \frac{\lambda}{\hbar c} - \frac{i\gamma E}{\hbar c \varepsilon} \right) \alpha_0 + \left( \delta - \frac{i\gamma mc^2}{\hbar c \varepsilon} \right) \beta_0 = 0,$$

$$\left( \delta + \frac{i\gamma mc^2}{\hbar c \varepsilon} \right) \alpha_0 + \left( \frac{\lambda}{\hbar c} + \frac{i\gamma E}{\hbar c \varepsilon} \right) \beta_0 = 0.$$

In order that $|\alpha_0|^2 + |\beta_0|^2 \neq 0$, the determinant of the coefficients must vanish. This gives $\delta^2 = (\lambda/\hbar c)^2 - \gamma^2$, which is analogous to the Dirac-Coulomb result. There are two possible values for $\delta$. Taking the positive one (for obvious reasons) we get

$$\frac{\beta_0}{\alpha_0} = -\frac{(\lambda/\hbar c) - (i\gamma E/\hbar c\varepsilon)}{\delta - i(\gamma mc^2/\hbar c\varepsilon)} = -\frac{\delta + i(\gamma mc^2/\hbar c\varepsilon)}{(\lambda/\hbar c) + (i\gamma E/\hbar c\varepsilon)}. \quad (55)$$

Returning to (54), we obtain the following recursion relations:

$$\left( q + \delta - \frac{i\gamma mc^2}{\hbar c \varepsilon} \right) \beta_q = -\frac{i}{2\hbar c} \alpha_{q-1} - \left( \frac{\lambda}{\hbar c} - \frac{i\gamma E}{\hbar c \varepsilon} \right) \alpha_q + \frac{1}{2} \beta_{q-1}, \quad (56a)$$

$$\left( q + \delta + \frac{i\gamma mc^2}{\hbar c \varepsilon} \right) \alpha_q = \frac{1}{2} \alpha_{q-1} + \frac{i}{2\hbar c} \beta_{q-1} - \left( \frac{\lambda}{\hbar c} + \frac{i\gamma E}{\hbar c \varepsilon} \right) \beta_q. \quad (56b)$$

Now, by some tedious (and uninteresting) algebraic manipulations similar to those in Greiner[14], we get: ($\bar{a}$ equals $a$ conjugate)

$$\frac{\beta_q}{\alpha_q} = \frac{q+\overline{K}}{i(q+K)}, \quad q \geq 1, \quad K = \delta + i\frac{\lambda}{\hbar c} - i\frac{\gamma}{\hbar c\varepsilon}(mc^2 - iE), \tag{57}$$

$$\alpha_q = \frac{q-1-n'}{q(q+2\delta)}\left(\frac{q+K}{q-1+K}\right)\alpha_{q-1}, \quad q \geq 2, \quad n' = \frac{\gamma E}{\hbar c\varepsilon} - \delta. \tag{58}$$

Iterating equation (58) gives:

$$\alpha_q = \frac{(q-1-n')(q-2-n')\cdots(1-n')}{q!(q+2\delta)(q-1+2\delta)\cdots(2+2\delta)}\left(\frac{q+K}{1+K}\right)\alpha_1. \tag{59}$$

We now must express $\alpha_1$ in terms of $\alpha_0$. To do this, set $q=1$ in equations (56a) and (56b) and get:

$$\left(i(1+\delta) + \frac{\gamma mc^2}{\hbar c\varepsilon}\right)\beta_1 = \frac{1}{2}\alpha_0 + \frac{i}{2\hbar c}\beta_0 - \left(\frac{i\lambda}{\hbar c} + \frac{\gamma E}{\hbar c\varepsilon}\right)\alpha_1, \tag{60a}$$

$$\left((1+\delta) + \frac{i\gamma mc^2}{\hbar c\varepsilon}\right)\alpha_1 = \frac{1}{2}\alpha_0 + \frac{i}{2\hbar c}\beta_0 - \left(\frac{\lambda}{\hbar c} + \frac{i\gamma E}{\hbar c\varepsilon}\right)\beta_1. \tag{60b}$$

Use one of these equations to eliminate $\beta_1$, so that the other can be used to determine $\alpha_1$ in terms of $\alpha_0$ and $\beta_0$. Since $\beta_0$ is proportional to $\alpha_0$ from equation (55), we obtain

$$\alpha_1 = \left(\frac{1+K}{1+2\delta}\right)\left(\frac{K - 2(i\lambda/\hbar c)}{\delta - i(\gamma mc^2/\hbar c\varepsilon)}\right)\frac{\alpha_0}{2} \Rightarrow$$

$$\alpha_q = \frac{(q-1-n')\cdots(1-n')}{q!(q+2\delta)\cdots(2+2\delta)}\left(\frac{q+K}{1+2\delta}\right)\left(\frac{K - 2(i\lambda/\hbar c)}{\delta - i(\gamma mc^2/\hbar c\varepsilon)}\right)\frac{\alpha_0}{2}. \tag{61}$$

Recall that the confluent hypergeometric function is defined by (see Gradshteyn and Ryzhik[17])

$${}_1F_1(a;b;x) = \sum_{n=0}^{\infty} \frac{(a)_n}{(b)_n}\frac{x^n}{n!}, \quad (a)_n = a(a+1)\cdots(a+n-1) = \frac{\Gamma(a+n)}{\Gamma(a)}. \tag{62}$$

We can write:

$$(q+2\delta)(q-1+2\delta)\cdots(2+2\delta)(1+2\delta)$$
$$= (1+2\delta)(1+2\delta+1)\cdots(1+2\delta+q-1) = (1+2\delta)_q,$$

and, using the decomposition $q+K = q-n' +[(n'+K)/(-n')](-n')$, in the numerator of equation (61), we get:

$$\begin{aligned}(q-1-n')(q-2-n')\cdots(1-n')(q+K) \\ = [(n'+K)/(-n')](-n')(-n'+1)\cdots(-n'+q-1) \\ = (1-n')_q + [(n'+K)/(-n')](-n')_q.\end{aligned}$$

We now have that

$$\begin{aligned}\phi_1(y) &= y^\delta \left(\frac{K-2(i\lambda/\hbar c)}{\delta - i(\gamma mc^2/\hbar c\varepsilon)}\right)\frac{\alpha_0}{2}\sum_{q=0}^{\infty}\frac{\left[(1-n')_q + [(n'+K)/(-n')](-n')_q\right]}{q!(1+2\delta)_q}y^q \\ &= y^\delta \left(\frac{K-2(i\lambda/\hbar c)}{\delta - i(\gamma mc^2/\hbar c\varepsilon)}\right)\frac{\alpha_0}{2}\left\{{}_1F_1(1-n';1+2\delta;y) + \left[\frac{(n'+K)}{(-n')}\right]{}_1F_1(-n';1+2\delta;y)\right\}.\end{aligned}$$ (63)

In a similar fashion, we can solve for the $\beta$ coefficients to get:

$$\beta_q = i\frac{(q+\bar{K})(q-1-n')\cdots(1-n')}{q!(q+2\delta)\cdots(2+2\delta)(1+2\delta)}\left(\frac{K-2(i\lambda/\hbar c)}{\lambda/\hbar c - i(\gamma E/\hbar c\varepsilon)}\right)\frac{\beta_0}{2}. \qquad (64)$$

This leads to

$$\phi_2(y) = y^\delta \left(\frac{K-2(i\lambda/\hbar c)}{\lambda/\hbar c - i(\gamma E/\hbar c\varepsilon)}\right)\frac{i\beta_0}{2}\left\{{}_1F_1(1-n';1+2\delta;y) + \left[\frac{(n'+\bar{K})}{(-n')}\right]{}_1F_1(-n';1+2\delta;y)\right\}. \qquad (65)$$

We can now obtain the eigenfunctions using

$$\Phi = \begin{bmatrix}\Phi_1 \\ \Phi_2\end{bmatrix} = \begin{bmatrix}\varsigma a(r) \\ \eta b(r)\end{bmatrix} = \begin{bmatrix}\begin{pmatrix}\eta_1 \\ -\eta_2\end{pmatrix}a(r) \\ \begin{pmatrix}\eta_1 \\ \eta_2\end{pmatrix}b(r)\end{bmatrix}, \qquad (66)$$

where

$$a(r) = (2\varepsilon/y)(mc^2 - E)^{1/2} e^{-y/2} (\phi_1 - \phi_2)$$

$$= \varepsilon y^{\delta-1} (mc^2 - E)^{1/2} e^{-y/2} \left( \frac{K - 2(i\lambda/\hbar c)}{\delta - i(\gamma mc^2/\hbar c \varepsilon)} \right) \alpha_0 \tag{67}$$

$$\times \left\{ (1+i) \left[ {}_1F_1(1-n'; 1+2\delta; y) - {}_1F_1(-n'; 1+2\delta; y) \right] + \left[ \frac{(K+i\bar{K})}{(-n')} \right] {}_1F_1(-n'; 1+2\delta; y) \right\},$$

and

$$b(r) = (2\varepsilon/y)(mc^2 + E)^{1/2} e^{-y/2} (\phi_1 + \phi_2)$$

$$= \varepsilon y^{\delta-1} (mc^2 + E)^{1/2} e^{-y/2} \left( \frac{K - 2(i\lambda/\hbar c)}{\delta - i(\gamma mc^2/\hbar c \varepsilon)} \right) \alpha_0 \tag{68}$$

$$\times \left\{ (1-i) \left[ {}_1F_1(1-n'; 1+2\delta; y) - {}_1F_1(-n'; 1+2\delta; y) \right] + \left[ \frac{(K-i\bar{K})}{(-n')} \right] {}_1F_1(-n'; 1+2\delta; y) \right\}.$$

Thus, we see that the solutions of the radial equations are linear combinations of the same types of confluent hypergeometric functions that occur as eigenfunctions of the Dirac-Coulomb problem. However, our parameters ($n'$ and $\delta$) are different from corresponding ones that appear in the Dirac-Coulomb case. Furthermore, our coefficients are complex numbers.

To obtain the eigenvalues $E$, we follow standard procedures and assume that $n'$ is an integer so that the series reduce to a family of polynomials. This gives

$$E_{n'} = \pm mc^2 \left[ 1 + \frac{\gamma^2}{n' + \sqrt{(\lambda/\hbar c)^2 - \gamma^2}} \right]^{-1/2}. \tag{69}$$

We see that the energy eigenvalues are of the same form as in the Dirac-Coulomb case. The only difference is that the quantity $j + \tfrac{1}{2}$ in the Dirac-Coulomb problem is replaced by $\lambda/\hbar c$ (where $j + \tfrac{1}{2}$ is the total angular momentum quantum number). We will discuss the implications of our solution on the foundations of QED after a brief look at the angular equations.

**The Angular Equations**

We now investigate equation (49). Using the matrix (47) for $\Omega$, the angular equations become, upon reinserting the appropriate factors of $\hbar$, and using the spinor relations: ($\tilde{m} = m/\hbar, z = e|\mathbf{\mu}_I|/\hbar c$)

$$\left( \frac{d}{d\theta} + \tfrac{1}{2}\cot\theta \right) \eta_2 + (\tilde{m}\csc\theta - z\sin^2\theta)\eta_2 = -\frac{\lambda}{\hbar c}\eta_1 \tag{70a}$$

$$\left(\frac{d}{d\theta} + \tfrac{1}{2}\cot\theta\right)\eta_1 - \left(\tilde{m}\csc\theta - z\sin^2\theta\right)\eta_1 = \frac{\lambda}{\hbar c}\eta_2. \tag{70b}$$

Making the change of variable $x = \cos\theta$ transforms (70) to

$$\left[(1-x^2)^{1/2}\frac{d}{dx} - \frac{(\tilde{m} + \tfrac{1}{2}x)}{(1-x^2)^{1/2}} + z(1-x^2)\right]\eta_2 = \frac{\lambda}{\hbar c}\eta_1, \tag{71a}$$

$$\left[(1-x^2)^{1/2}\frac{d}{dx} + \frac{(\tilde{m} + \tfrac{1}{2}x)}{(1-x^2)^{1/2}} - z(1-x^2)\right]\eta_1 = -\frac{\lambda}{\hbar c}\eta_2. \tag{71b}$$

These are generalizations of equations that lead to the associated Legendre equations. In order to see the differences, use (71a) to solve for $\eta_1$ and put this in (71b) to obtain an equation for $\eta_2$ (the $\eta_1$ equation is identical so we drop the subscripts and $\tilde{\lambda} = \lambda/\hbar c$):

$$\begin{aligned}\left\{(1-x^2)\eta'' - x\eta' + \left[\tilde{\lambda}^2 - \tfrac{1}{4} - \frac{\tilde{m}^2 - \tilde{m}x + \tfrac{1}{4}}{(1-x^2)}\right]\eta\right.\\ \left.+ \left[2z(\tilde{m}+x)(1-x^2)^{1/2} - z^2(1-x^2)^2\right]\eta\right\} = 0.\end{aligned} \tag{72}$$

The term in braces is close to the (general) form of Legendre's differential equation, but differs by a factor of two in the $\eta'$ term (see Gradshteyn and Ryzhik[17]). The deviation from Legendre's equation reflects the lack of spherical symmetry, caused in part by the magnetic moment potential, and the separation of variables method. Thus, the solutions of (72) will be generalizations of the Legendre functions. We are currently unable to construct an exact analytic solution of equation (72). Interestingly, it is the square root term that prevents the use of standard (analytic) solution methods.

**Problems**

We can now identify the problems that cloud our complete understanding of the Dirac equation and its relation to the hydrogen spectrum.

**Physical**

The basic physical problem is to construct a complete solution for the angular eigenvalue problem for the Dirac equation with the Coulomb and magnetic dipole interaction (equation (72)).

**Mathematical**

A basic mathematical problem is to prove or disprove that perturbation theory can (or cannot) be applied to equation (25a) and the $\mathbf{A}^2$ term equation (28) (using (27)).

Since equation (72) is not one of the standard forms, we must use approximation methods to solve it. It appears to require some new ideas for its analytic solution. This would be preferred since our separation of variables method does not allow us to directly explore the questions posed. For example, we know that the radial equation has the same singularity properties at the origin as the Dirac-Coulomb case, but we still cannot say that perturbation analysis of the $\mathbf{A}^2$ term is justified.

**Conclusion**

In this paper we have shown that the full (minimal coupling) Dirac equation can be analytically separated (diagonalized) into particle and antiparticle components without transforming the wave functions, as is done by the Foldy-Wouthuysen method. This diagonalization reveals the nonlocal time behavior of the particle-antiparticle relationship. We have shown that a more physically reasonable interpretation of the zitterbewegung, and the result that a velocity measurement (of a Dirac particle) at any instant in time is $\pm c$, are reflections of the fact that the Dirac equation makes a spatially extended particle appear as a point in the present by forcing it to oscillate between the past and future at speed $c$.

We have also shown that one of the difficult issues facing attempts to completely understand the Dirac problem for full coupling is the singular nature of the $\mathbf{A}^2$ term. This term is small in all but s- states, where it diverges when treated as a perturbation. If we introduce a cutoff, the contribution is of order $\gamma^7$ so one might be inclined to dismiss the term (as is traditionally done). However, this term appears to be more singular than the Coulomb potential, so that perturbation analysis, and indeed, the whole eigenvalue approach may be called into doubt. On the other hand, this is not completely clear since the $\sin^2\theta$ term vanishes on the spin axis and could strongly modify the singular nature of this term. This problem must be solved in order to determine the exact extent that the Dirac equation contributes to spectrum of hydrogen. If the problems posed in this paper can be solved in the positive, it would appear that the correct approach for s-state (hyperfine) splitting gives the same results as those obtained from using the Pauli equation. Furthermore, equation (24b) introduces a natural cutoff, which removes the conceptual difficulty of a point magnetic dipole interaction as implied by use of the delta term in the Pauli equation. This also suggests that, the use of cutoffs in QED, are already justified by the eigenvalue analysis that supports it.

Using a different method, we are able to effect separation of variables for full coupling and solve the radial equation. Since the behavior of the radial equation at the origin is the same as in the Dirac-Coulomb case, we can say that the $\mathbf{A}$ term does not increase the singular nature of the radial equation. We have not been able to solve the angular equation. Although we strongly believe that contribution is small, we still cannot say how much the Dirac equation contributes to the hydrogen spectrum.

## ACKNOWLEDGMENT

The authors would like to acknowledge important comments and encouragement from Professor David Finkelstein.